\documentclass[twocolumn,tighten]{aastex63}
\usepackage{graphicx}
\usepackage{gensymb}

\shorttitle{Optical Detection of the SN1941C Remnant}
\shortauthors{Fesen and Weil}

\newcommand{\unit}[1]{\, {\rm #1}}

\begin{document}
\title{Detection of Late-Time Optical Emission from SN 1941C in NGC 4136 } 

\author[0000-0003-3829-2056]{Robert A.\ Fesen}
\affil{6127 Wilder Lab, Department of Physics and Astronomy, Dartmouth
                 College, Hanover, NH 03755 USA}

\author[00000-0002-4471-9960]{Kathryn E.\ Weil}
\affil{6127 Wilder Lab, Department of Physics and Astronomy, Dartmouth
                 College, Hanover, NH 03755 USA}
\affil{Smithsonian Astrophysical Observatory, 60 Garden Street, Cambridge, MA 02138, USA}


\begin{abstract}

We report the detection of broad, high-velocity oxygen emission lines from the
site of SN~1941C nearly eight decades after outburst, making it the oldest
optically detected historical core-collapse supernova (CCSN) and one of the youngest
core-collapse supernova remnants with a well determined age. In contrast to the strongly
blueshifted emission line profiles observed for other late-time CCSNe
thought to be due to internal dust extinction of far-side hemisphere of expanding ejecta, SN~1941C's
spectrum exhibits stronger redshifted than blueshifted emissions of [O~I] 6300,
6364 \AA, [O~II] 7319, 7330 \AA, and [O~III] 4959, 5007 \AA. These oxygen emissions exhibit
rest frame expansion velocities from $-2200$ to $+4400$ $\unit{km} \unit{ s}^{-1}$. No other
significant broad line emissions were detected including H$\alpha$ emission. We
discuss possible causes for this unusual spectrum and compare SN~1941C's
optical and X-ray luminosities to other evolved CCSNe.

\end{abstract}
\bigskip


\section{Introduction}

The transition of supernovae (SNe) into supernova remnants (SNRs) is an
important yet poorly studied evolutionary phase \citep{Mili2017}.  Since SNe
typically fade several magnitudes a year after outburst, detections of SNe more
than one or two decades after maximum light in the optical, radio, or X-rays
are rare, with nearly all late-time detections associated with core-collapse
SNe (CCSNe). This has made it difficult to establish links between the
properties of the few young Galactic SNRs with ages between 100 and 2000 yr,
which can be studied in far greater detail than their extragalactic
counterparts, with the variety of observed SN types and subclasses.

The fact that some CCSNe emitted significant emission a decade or more after
outburst was only realized in the late 1980's when SN~1957D in M83 and SN~1980K
in NGC~6949 were detected optically \citep{Turatto1989,Long1989,Fesen1990}.  In
those few cases where late-time SN emissions are observed, the most common
interpretation is an interaction of the SN's outwardly expanding forward shock
and clumpy ejecta with dense circumstellar material (CSM)
\citep{CF03,CF06,CF17,Mili2012}.  However, other late-time energy sources such
as pulsars, accretion around black holes, and magnetars have also been
suggested \citep{CF94,Woosley2010,Patnaude2011,Mili2018}.

Late-time radio and X-ray emissions from shocked progenitor wind material are
expected to decrease at a rate of $t^{-s}$ where s = 1 to 4
\citep{Weiler2002,Immler2005,Stockdale2006}, consistent with the few detections
reported. In contrast, late-time optical emissions are likely generated by
reverse shock-heating of dense ejecta due to the presence of a dense wind or
CSM relatively close to the progenitor star \citep{Mili2012,Black2017}. Dust
forming in the ejecta or in a cool dense shell of shocked SN ejecta (CDS)
located between the forward CSM shock and the reverse shock front can lead to
the commonly observed strongly blueshifted optical emission lines reported at
late-times
\citep{Fesen1990,Fesen1993,McCray1993,CF94,Andrews2010,Andrews2011,Mili2012,Bevan2017}
due to absorption of the redshifted, far- side of the expanding SN shell.
Consequently, optical emission lines most often show strong blue/red
asymmetries in lines such as H$\alpha$ and [\ion{O}{1}] 6300, 6364 \AA \ and
[\ion{O}{3}] 4959, 5007 \AA. 

Of the handful of CCSNe with reported late-time optical, X-ray, and radio
emissions decades after outburst, the majority are  either SN~II-L or SN~IIb
events.  These include SN~1970G in M101, SN~1979C in M100, SN~1980K in NGC~6946
and SN~1993J in M81. Such late-time emissions suggest significant CSM
immediately local to the SN \citep{Mili2012}, a property not commonly present
for SNe~IIP, although there are exceptions (SN 2004et: \citealt{Long2019}).
While a handful of extragalactic SNRs are have been found to be young and
ejecta-dominated, such as the luminous O-rich SNR in NGC~4449
\citep{Kirshner1980,Blair1983,Long2012} and B174a in M83 \citep{Blair2015},
they are not associated with specific historical SNe and hence their ages and
SN subtypes are not well constrained.  

At present, the oldest recorded SN exhibiting any sort of late-time emission is
SN~1923A in M83.  Although detected in the radio by several groups
\citep{Cowan1994,Eck1998,Stockdale2006}, no identifiable X-ray or optical
emission has been found \citep{Blair2004}.  Until now, the oldest SN exhibiting
late-time optical and X-ray emissions is SN~1957D also in M83
\citep{Long2012,Mili2012} but, as is common with such old historical SN, its
subtype is unfortunately unknown \citep{Long1989}. 

Here we present the discovery of late-time optical emissions from an even older
object, namely SN~1941C. It was discovered by R.B.\ Jones in the southeastern
outskirts of the nearly face-on ($i$ = 25${\degree}$; \citealt{Fridman2005})
SBc spiral galaxy NGC~4136 using Harvard College photographic plates
\citep{Jones1941}.  Its apparent brightness between 1941 April 16 thru 22  was
between 16.8 and 17.1 mag \citep{Jones1941}.  Although there are no published
follow-up photometric or spectral observations, \citet{Adams1941} reported that
a spectrum taken by M.\ Humason and R.\ Minkowski indicated it was a Type II event.

\begin{figure*}[htp]
\includegraphics[width=0.3045\textwidth]{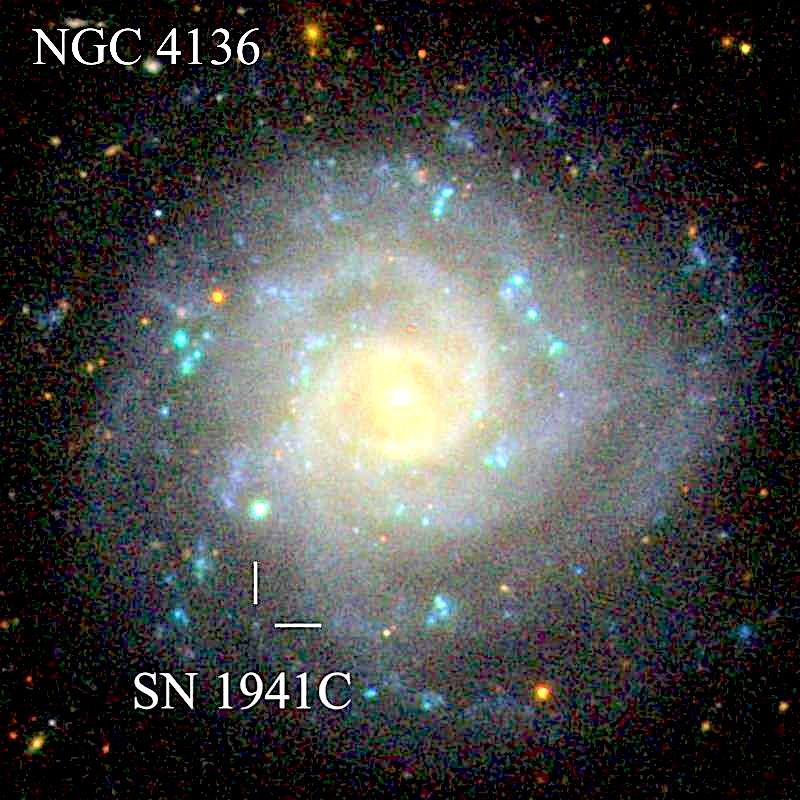}\hfill
\includegraphics[width=0.341\textwidth]{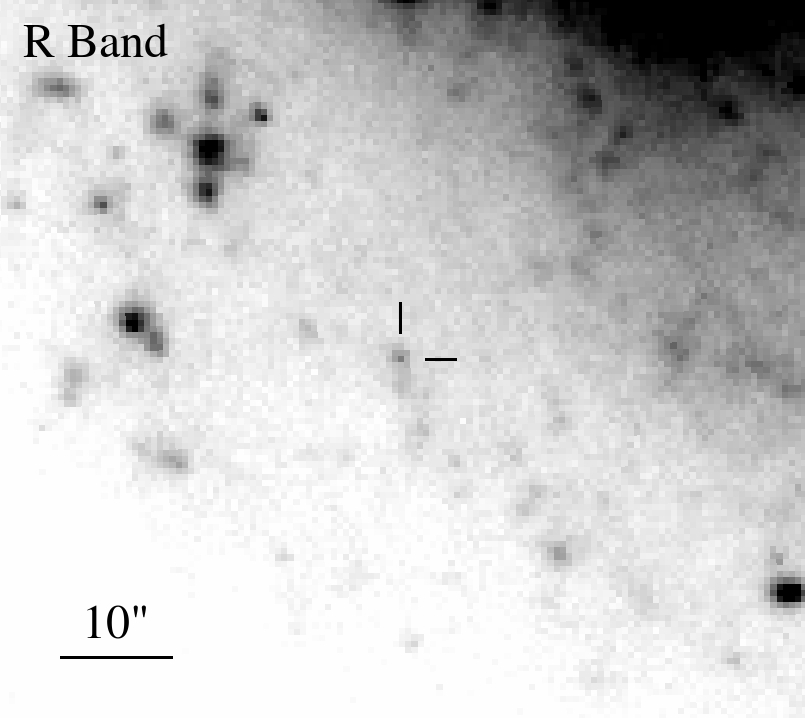}\hfill
\includegraphics[width=0.341\textwidth]{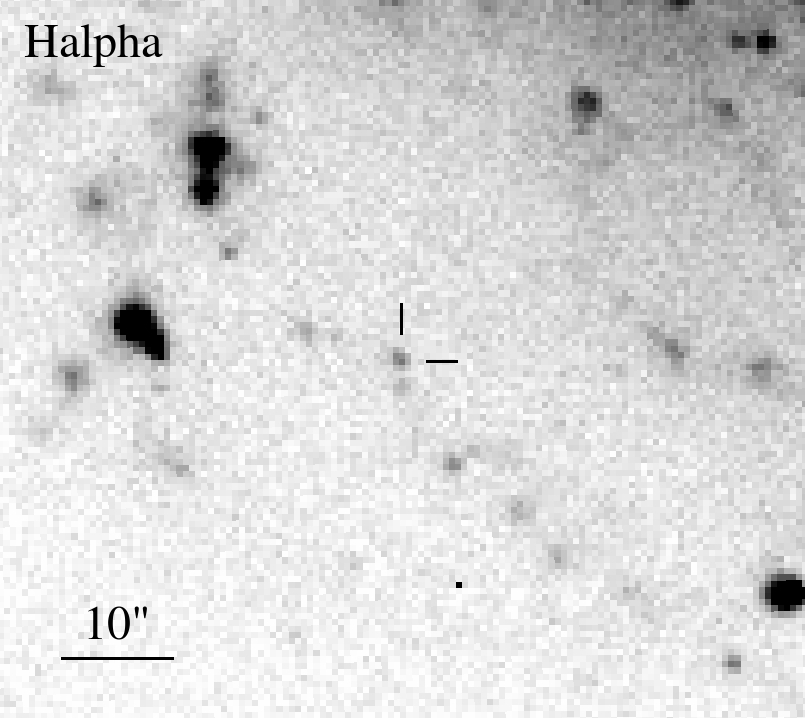}
\caption{Left panel: Color SDSS DR14 image of NGC~4136 
         (credit: C. Seligman) with the location of SN~1941C marked.
         Center and right panels: MDM R band and H$\alpha$ images of the SN~1941C site. }
\label{SDSS_image}
\end{figure*}

\begin{figure*}[htp]
\centering
\includegraphics[angle=0,width=17.0cm]{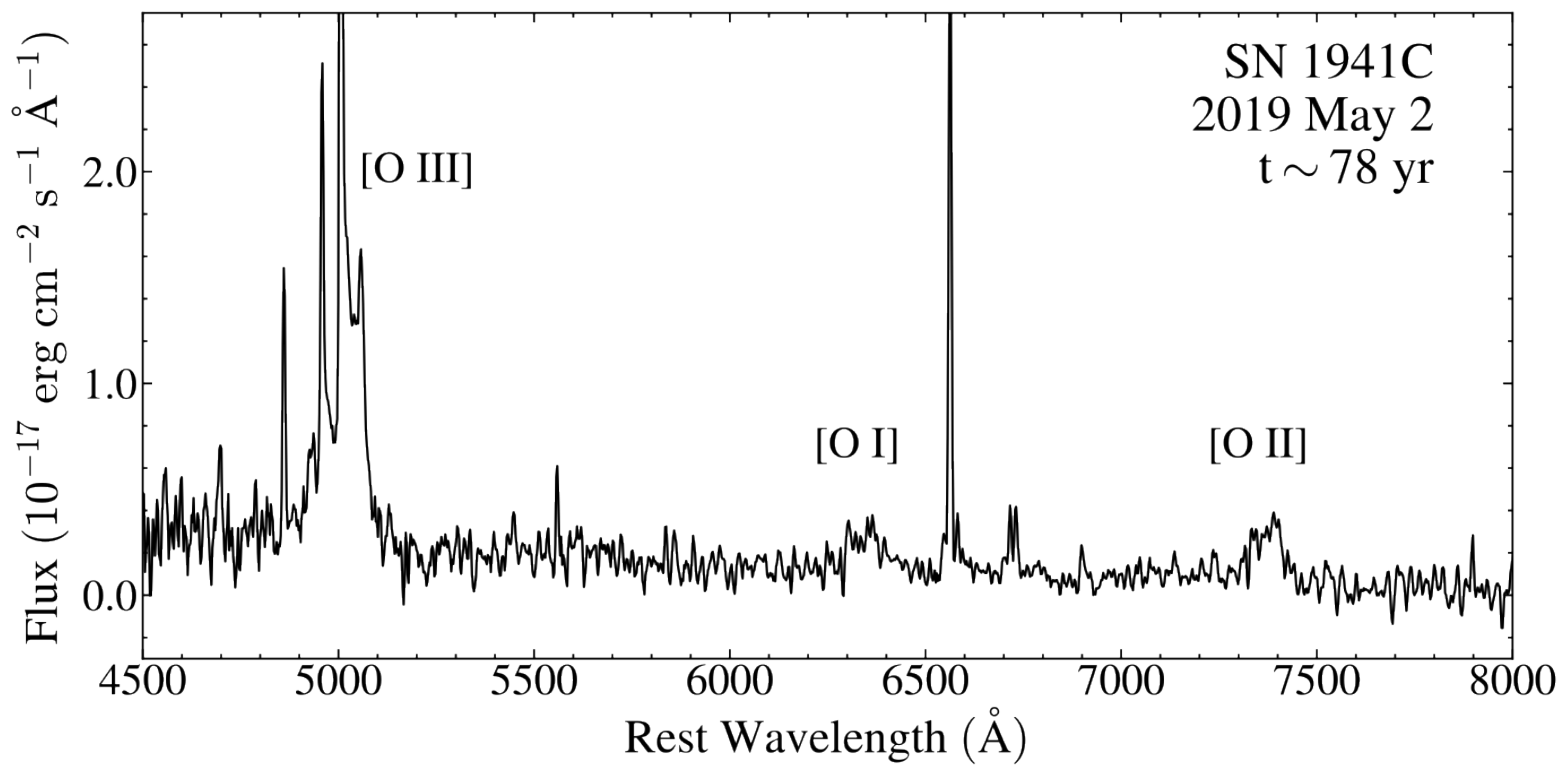} \\
\includegraphics[angle=0,width=17.0cm]{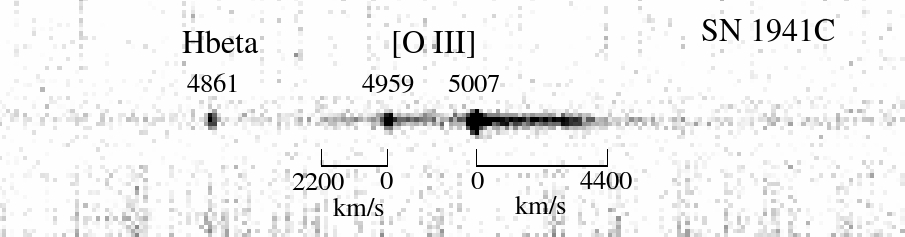}
\caption{Upper Panel: Low dispersion spectra of SN~1941C.
         Lower Panel: A 2D image of the background subtracted SN 1941C spectrum
         around the [\ion{O}{3}] 4959, 5007 \AA \  emission lines showing redshifted emission
         out to 4400 km s$^{-1}$ and fainter blueshifted emission to $-2200$ km s$^{-1}$.}
\label{spectra}
\end{figure*}

\begin{figure}[t]
\centering
\includegraphics[angle=0,width=0.8\columnwidth]{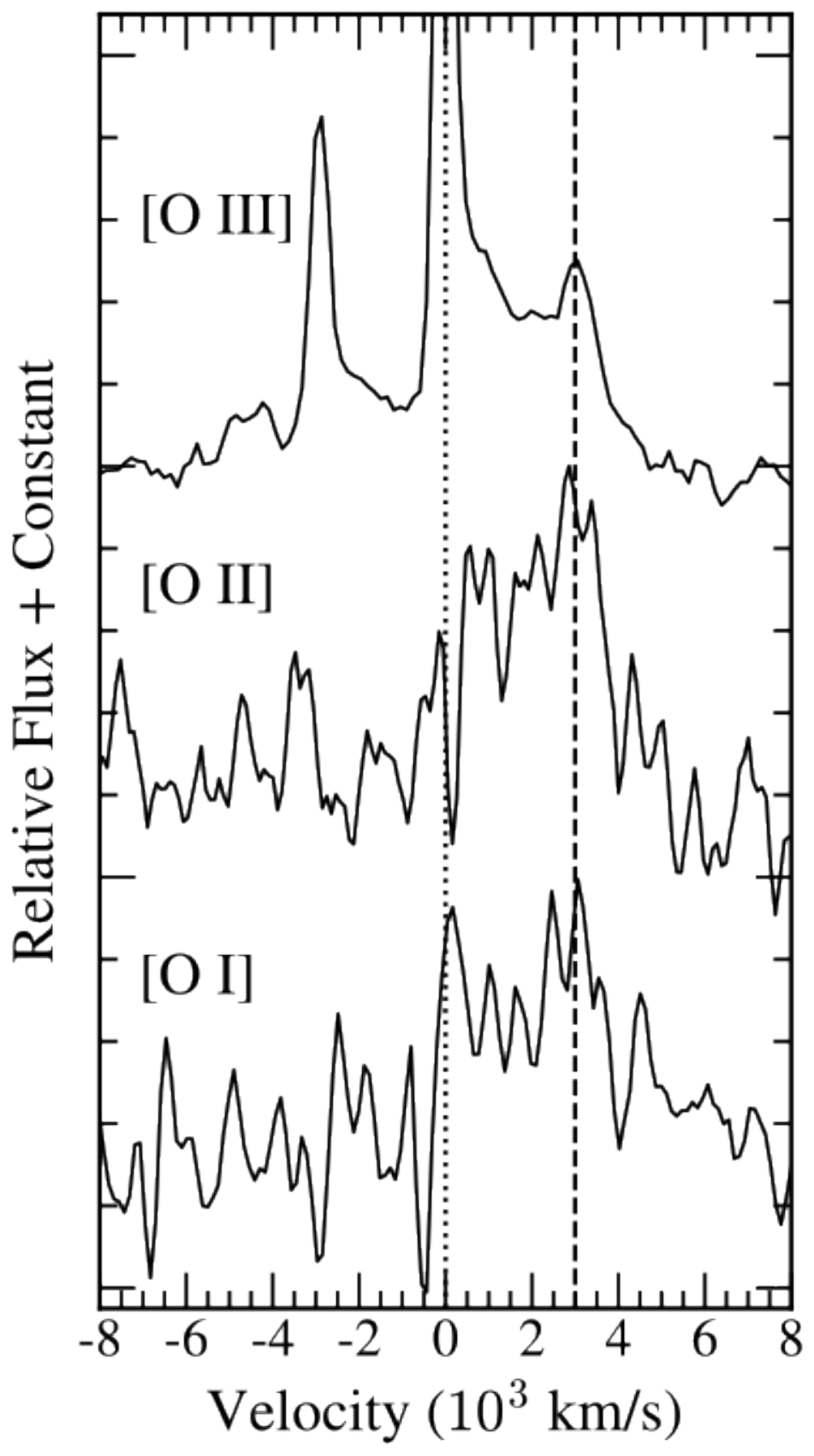}
\caption{Plot of forbidden oxygen line emissions relative to expansion velocity.
The vertical dotted line marks the location of the zero rest frame 
expansion velocity for [\ion{O}{1}] 6300 \AA, [\ion{O}{2}] 7325 \AA, 
and [\ion{O}{1}] 5007 \AA \ line emissions. 
The vertical dashed line marks the emission peak seen at 3000 km s${^-1}$ in all three
emission features. } 
\label{velocity_plot}
\end{figure}

Estimated distances to NGC~4136 vary from $8 \pm 2$ Mpc \citep{Gusev2003}, 10.5
Mpc \citep{Bott1984}, and 9.7 Mpc \citep{Tully1988}. Here we will adopt a
distance of 9.7 Mpc. The galaxy has only a small amount of foreground Galactic
extinction of A$_{\rm V}$ = 0.05 mag \citep{SF2011}.

Although radio searches by \citet{Bruyn1973} and \citet{Eck2002} for associated
radio emission from historical extragalactic SNe did not survey NGC~4136, an
upper limit of 4 mJy at 2695 MHz was reported by \citet{Brown1978}.  However, a
search of archival {\sl Chandra} X-ray  data covering locations of historical SN
older than SN~1970G in parent galaxies closer than 15 Mpc revealed an X-ray
point source at the location of SN~1941C with a luminosity in 2002 of $\approx
5 \times 10^{37}$ erg s$^{-1}$ \citep{Soria2008}\footnote{X-ray emission from
SN~1941C can be seen in the Chandra images of NGC~4136 published by
\citealt{Roberts2004}.}. 

The detection of significant X-ray flux from SN~1941C some six decades after
outburst with an X-ray flux comparable to other old CCSNe \citep{Dwark2012}
encouraged us to investigate possible late-time optical emission from this Type
II SN.  Our imaging and spectral data are described in $\S$2, with results 
given in $\S$3, with our discussion and conclusions in $\S$4.

\section{Observations}

A series of 600 s R band and H$\alpha$ images of the SN~1941C site in  NGC~4136
were taken on 2019 May 2 with the 2.4m Hiltner telescope at MDM Observatory
using the Ohio State Multi-Object Spectrograph (OSMOS,  \citealt{Martini2011}).
These were followed by a series of low-dispersion, long-slit OSMOS spectra.
Using a 1.4 arcsec N-S slit and a red VPH grism (R = 3000), providing a
wavelength coverage $4000
- 8600$ \AA \ and a spectral resolution of 7 \AA, we obtained four 3000 s to
  4000 s exposures which were then summed into a single 15,000 s exposure.   We
also obtained a 4000 s exposure using this same setup on the young, O-rich SNR
in NGC~4449 for spectral and flux comparison purposes.

Standard pipeline data reduction of MDM images and spectra using AstroPy and 
PYRAF\footnote{PYRAF is a product of the Space Telescope Science Institute,
which is operated by AURA for NASA.} were performed.  Spectra were reduced using
the software L.A.\ Cosmic \citep{vanDokkum2001} to remove
cosmic rays.  Spectra were calibrated using an Ar lamp and spectroscopic
standard stars \citep{Oke74,Massey90}.


\section{Results}

The site of SN~1941C in NGC~4136 is shown in Figure~\ref{SDSS_image}, which
presents a color Sloan DSS image of the galaxy (left panel) along with MDM R
and H$\alpha$ images (center \& right panels). 
We found a point-like source in the R band image at the northern tip
of a small ($2'' \times 4''$) star cluster and coincident with a faint H~II
region along the southeastern edge of the galaxy with J2000 coordinates of RA =
12:09:21.01, Dec = +29:54:32.5.  This source is some $67\farcs1$ South
and $42\farcs5$ East of the galaxy's nucleus, consistent with the reported
offsets of 67\arcsec\ South and 44\arcsec\ East for SN 1941C. 

Figure~\ref{spectra} shows an optical spectrum of this point source (upper
panel) taken 78 yr after SN outburst covering the wavelength region 4500 to
8000 \AA, corrected for the 609 km s$^{-1}$ (z = 0.00203) redshift of NGC~4136.
Broad emission lines of [\ion{O}{3}] 4959, 5007 \AA, [\ion{O}{1}] 6300, 6364
\AA, and [\ion{O}{2}] 7319, 7330 \AA \ can be seen, along with narrow,
unresolved H~II region emission features.  Measured fluxes of the  broad
[\ion{O}{1}],  [\ion{O}{2}], and [\ion{O}{3}] emissions are $1.4 \pm 0.2$,
$1.80 \pm 0.25$, and $9.1 \pm 0.2$ $\times 10^{-16}$ erg s$^{-1}$ cm$^{-2}$,
respectively.  No broad H$\alpha$ emission was detected down to a flux level of
$4 \times 10^{-17}$ erg s$^{-1}$ cm$^{-2}$.

The spectrum shows redshifted [\ion{O}{3}] emission to be brightest, extending
to 5080 \AA\ or +4400 km s$^{-1}$ with a hint of possible higher velocities up
to $\simeq$ +5000 km s$^{-1}$. Blueshifted emission is also present but is much
fainter and extends to much lower velocities.  We observe the [\ion{O}{3}] 4959
\AA \ line  emission out only to 4922 \AA \ or $-2200$ km s$^{-1}$ (see
Fig.~\ref{spectra}, lower panel).  

Figure~\ref{velocity_plot} shows a comparison of emission profiles for all
three forbidden oxygen emission lines in terms of expansion velocity. A dashed
line marks the +3000 km s$^{-1}$ emission feature.  Although the low S/N of the
data prevents confirmation of the full velocity range seen in the [\ion{O}{3}]
lines in the [\ion{O}{1}] and [\ion{O}{2}] lines, it is clear that the
SN~1941C's redshifted, hemisphere O-rich ejecta constitutes its brightest
late-time optical emission.

 Although
the 3000 km s$^{-1}$ emission feature is strongest in [\ion{O}{3}] in terms of flux, the peak's
relative strength appears strongest in the [\ion{O}{2}] 7320, 7330 \AA \ emission.  This might
indicate the presence of significant dust in the O-rich ejecta, like that
commonly seen in late-time spectra of other CCSNe.


\section{Discussion}

The site of SN~1941C is coincident with an H~II region, consistent with it being a Type
II SN and thus a CCSN just as Humason and  Minkowski's 1941 spectrum
suggested. Unfortunately, no archival {\sl Hubble Space Telescope} images are
available of the SN~1941C site in NGC~4136 which might inform us about the
stellar neighborhood of this CCSN leading to possible estimates of its
progenitor mass.

Unlike the other nearly as old historical SN, namely the 62 yr old SN~1957D in
M83 which is currently near the limit of optical detectability
\citep{Long2012,Mili2012}, SN~1941C's late-time 78 yr (t $\simeq$ +28 500 day)
optical emission is relatively bright.  While this is maybe not so surprising
given its observed 2002 X-ray luminosity \citep{Soria2008}, SN~1941C's
late-time optical spectrum is remarkable for displaying such bright redshifted
emission compared to its much fainter blueshifted emission.  This it is in contrast to
nearly all late-time CCSNe where the SN's far-side hemisphere of ejecta emissions
are strongly affected by dust which produces the observed strongly blueshifted
emission profiles.  Since we observed no broad H$\alpha$ emission which would
be the signature of dense CSM, we conclude that any dust in the SN~1941C
remnant is largely confined to the ejecta, like that inferred for other CCSN
exhibiting dust such as in SN~1979C \citep{Fesen1993,Mili2012}.

The presence of a bright emission feature at 3000 km s$^{-1}$ 
is similar to bright emission peaks off from line center 
seen in other late-time SN emissions, such as the SNR 4449-1 \citep{Mili2008}. 
Often attributed in the past to the presence of large clumps of O-rich ejecta, 
it is more likely due to the presence of a discrete ring(s) of ejecta such 
as observed in the well resolved young remnant, Cas~A \citep{Delaney2010,Mili2013}.
If correct, this suggests the formation of large expanding rings may be a common feature in
CCSN remnants and, in turn, signal the presence of large bubble structures 
possibly generated by post-SN heating due to radioactive element-rich ejecta. 

\begin{deluxetable*}{llccccccccl}
\tablecolumns{11}
\tablewidth{0pt}
\tablecaption{Expansion Velocities\tablenotemark{a} and Luminosities\tablenotemark{b} of Young Ejecta Dominated Core-Collapse SNRs}
\label{fluxes}
\tablehead{\colhead{Object}& \colhead{SN}   & \colhead{Host}   & \colhead{Distance} & \colhead{Age}  & \colhead{V$_{\rm exp}$} &
 \multicolumn{3}{c}{\underline{~~~Lum. (10$^{36}$ erg s$^{-1}$)~~~}}  & \colhead{Obs.\ Epochs} & \colhead{References} \\
 \colhead{ }    & \colhead{Type} & \colhead{Galaxy} & \colhead{(Mpc)}    & \colhead{(yr)} & \colhead{(km s$^{-1}$)}  &
 \colhead{[O I]} & \colhead{ [O III] }   & \colhead{X-ray } & \colhead{opt:X-ray} & \colhead{ } }
\startdata
Cassiopeia A\tablenotemark{c} &  IIb  & Milky Way  & 0.0034   & $\sim$350   & $-4500$ to +6500   & 0.5    &  2.3  & 27  & 1996:2012  & 1, 2    \\
SNR B12-174a     & \nodata & M83        &  4.6     & $\leq$100   & $\pm 5200$         & 6.3    &  2.5  & 7.3    & 2011:2012  & 3, 4    \\
 \bf{SN 1941C}   & II ~    & NGC 4136   &  9.7     &  ~ 78       & $-2200$ to +4400   & 1.7    &  11   & 50     & 2019:2002  & 5, 6    \\
SNR 4449-1       & \nodata & NGC 4449   &  3.9     & $\sim$70    & $\pm5700$          & 26     &  145  & 240    & 2019:2001  & 5, 7, 8, 9 \\
SN 1957D         &  II ~   & M83        &  4.6     &  ~ 62       & $\pm2700$          & 0.3    &  1.1  & 17     & 2011:2011  & 10, 11  \\
SN 1970G         &  IIL    & M101       &  7.4     &  ~ 49       & $-6000$ to +2200   & 3.0    &  10   & 41     & 2010:2011  & 11, 12  \\
SN 1979C         &  IIL    & M100       &  15      &  ~ 40       & $-6100$ to +5000   & 120    &  260  & 650    & 2016:2007  & 11, 12, 13 \\
SN 1980K         &  IIL    & NGC 6946   &  5.9     &  ~ 39       & $-4200$ to +2100   & 2.4    & 1.6   & 30     & 2010:2004  & 11, 14   \\
SN 1986J         &  IIn    & NGC 891    &   10     &  ~ 36       & $-3800$ to +2400   & 24     & 4.6   & 1800   & 2007:2003  & 11, 15   \\
SN 1986E         &  IIL    & NGC 4302   &   17     &  ~ 33       & $-4800$ to +2600   & 47     & 10    & \nodata& 1994:\nodata & 16      \\
SN 1993J         &  IIb    &  M81       &  3.6     &  ~ 26       & $-5800$ to +7000   & 1.6    & 15    & 200    & 2009:2008  & 11, 17    \\
SN 1996cr        &  IIn    & Circinus   &  3.8     &  ~ 23       & $\pm4300$          & 11     & 5.4   & 4000   & 2017:2007  & 11, 18, 19      \\
\enddata
\tablenotetext{a}{Velocities correspond to [\ion{O}{1}] or [\ion{O}{3}] emissions. }
\tablenotetext{b}{Optical [\ion{O}{1}] 6300 + 6364 \AA, [\ion{O}{3}] 4959 + 5007 \AA, and X-ray luminosities are calculated
                  assuming the distance listed for the host galaxy. See \citet{Dwark2012} for tabulated X-ray energy range (keV) used to calculate X-ray luminosities. }
\tablenotetext{c}{Values for Cas A are taken from \citet{Winkler2017} which include extinction corrections assuming A$_{\rm V}$ = 6.2.}
\tablenotetext{}{References: 1: \citet{Winkler2017};  2: \citet{Mili2013};
                             3: \citet{Long2014};  4: \citet{Blair2015};
                             5: this paper; 6: \citet{Soria2008};
                             7: \citet{Mili2008}; 8: \citet{Patnaude2003}; 
                             9: \citet{Summers2003};
                             10: \citet{Long2012}; 11: \citet{Mili2012};
                             12: \citet{Dittmann2014};
                             13: \citet{Patnaude2011}; 14: unpublished 2016 MMT spectrum;
                             14: \citet{Frid2008};
                             15: \citet{Houck2005};
                             16: \citet{Capp1995};
                             17: \citet{Chandra2009};
                             18: Patnaude et al.\ in prep; 
                             19: \citet{Bauer2008}
}

\end{deluxetable*}

Table~\ref{fluxes} shows a comparison of expansion velocities and relative optical
[\ion{O}{1}], [\ion{O}{3}] and X-ray luminosities for several historical,
extragalactic ejecta dominated SNe with known or estimated ages between 20 and
100 yr. We have also included the 350 yr old Galactic SNR, Cas~A for comparison.

This table shows that, excluding the extremely bright O-rich remnant in
NGC~4449, SN~1941C is quite luminous both in the optical and in X-rays for an object
older than 50 years. While there is no radio detection reported for this SN,
we expect it to be at least as bright as that of SN~1957D (0.6 mJy
at 6 cm in 1998; \citealt{Stockdale2006}) or SNR~B12-174a (0.2 mJy at 5 GHz in 2011;
\citealt{Blair2015}).

Table~\ref{fluxes} also reveals again how unusual SN~1941C's blue and redshifted expansion
velocities are, not in terms of absolute velocity, but in its unusual blue/red
asymmetry where the far hemisphere velocity is significantly brighter than that
seen for its near-side blueshifted  O-rich ejecta.  In this respect, SN~1941C may
represent a valuable transitional object where we are observing a true SNR
rather than just a late evolutionary emission stage of a recent SN.  Because
there is no appreciable H$\alpha$ emission present, the source of SN~1941C's
late-time emission would not appear to be due to current or recent CSM-ejecta
interaction, but more likely due to the progression of its reverse shock back
into the expanding ejecta.

In this respect, the remnant of SN~1941C is like that of Cas~A where its total
global emission is dominated by emissions from ejecta with only very faint
H$\alpha$ and [\ion{N}{2}] emissions \citep{Mili2008}. Moreover, Cas~A's
optical emission has radially changed in the roughly 70 year observing period (1950 to
present) where in the early 1950's only a few relatively faint optical
filaments along its northern limb were visible, whereas currently a nearly
complete and bright ring of ejecta emissions is seen \citep{Patnaude2014}.
Cas~A appears to also shared SN~1941C's asymmetry in its blue and red expansion velocities
(see Table~\ref{fluxes}).

It is possible that SN~1941C's O-rich ejecta are in a similar early stage of
being reverse shock-heated where relatively limited portions of the object's
expanding ejecta shell are visible leading to the blueshifted/redshifted
asymmetry observed.  Consequently, it may prove useful to monitor SN~1941C's
late-time optical and X-ray emissions to observe changes in both luminosity and
velocity.

The presence of an unknown amount of internal ejecta extinction, in both the
facing but especially the remnant's far-side redshifted hemisphere, makes mass
estimates based on observed emission fluxes quite uncertain. Nonetheless, a
comparison of the SN~1941C remnant's [\ion{O}{1}] and [\ion{O}{3}] luminosities
to the similar age O-rich SNe/young SNR seen in NGC 4449 (see
Table~\ref{fluxes}) which has an estimated oxygen mass $\sim 7.5 \times
10^{-2}$ M$_{\odot}$ \citep{Kirshner1980} suggests an oxygen mass $\sim
10^{-2}$ M$_{\odot}$.  Because larger expansion velocities hinted by our low
S/N spectra may indicate a more extensive O-rich expanding shell, this value
could be viewed as a lower limit.

Due to the relatively low quality and quantity of early 20th century photographic
spectra of SNe and the fact that the spectral derived classes of Type I and Type II
SNe was only established in 1941 \citep{Minkowski1941,Fili1997}, only about a
dozen of the 40 historical extragalactic SNe with ages greater than SN~1941C
have been classified or proposed as either possible or likely Type II events
\citep{Barbon2010}. However, some of these assigned SN types are
questionable.

Prior to the discovery of SN~1941C, \citet{Zwicky1964} listed just five SNe
(1936A, 1937A, 1940A, 1940B, and 1941A) as definitely or likely Type II SNe.
Subsequent SN lists by \citet{Kowal1971}, \citet{Sargent1974}, \citet{Maza1976}, and
\citet{Flin1979} increased this number by adding four more (1919A, 1926A,
1937F, and 1939C).  More recent listings such as the Asiago Supernova Catalogue
\citep{Barbon1989,Barbon1999,Barbon2010} added several more (1909A, 1917A, 1921A,
1923A, and 1940C) along with II-L and II-P light curve sub-classifications for
some.  

However, poor quality and/or the absence of spectral data have led to
differences in assigned SN types in various lists for some SNe prior to
1941C, suggesting the fidelity of some SN assigned types might not be high.  An
example is SN 1909A where its photographic light curve does not seem to fit
either that of a Type I or II \citep{Sandage1974,Patat1993}, resulting in a
variety of its assigned SN type in various catalogs:  peculiar, IV, peculiar,
II-P \citep{Sargent1974,Flin1979,Barbon1989,Barbon1999}.  

Another example is that of SN~1917A in NGC~6946. A 6 hr long slitless spectrum
obtained by Pease and Ritchey at least a month or so after maximum light
\citep{Ritchey1917,Shapley1917} was described as ``a strong continuous
spectrum, crossed by what appears to be a series of bright bands''.  A later
examination of this spectrum by Baade and Humason \citep{Baade1938} showed ``a
typical supernova spectrum with the wide $\lambda$4600 band as the dominant
feature'', a property common in post-max Type Ia spectra \citep{Black2016}.
Although this assessment was subsequently echoed by \citet{Mayall1948} in his
report on the spectrum of the Type II SN~1948B in the same galaxy, SN~1917A is
listed as a Type II in most recent SN catalogs
\citep{Barbon1999,Barbon2010,Lennarz2012,Guill2017}.   

Consequently, until emission from more historical CCSNe older than SN~1941C are
detected and studied across a range of wavelengths, our understanding of the
transition of CCSNe into core-collapse SNRs is limited.  The very small number
of detected young extragalactic CCSNe, unfortunately, does not offer much help
lluminating this evolutionary phase.  Only two Magellanic Clouds core-collapse
SNRs have estimated ages less than 2000 yr (namely, 1E0202-7219 and 0540-69.3)
and despite extensive SNR surveys in nearby galaxies including M31, M33, M83,
and NGC~6946 \citep{Long2017,Long2019} only two extragalactic CCSN remnants
outside of the Magellanic Clouds not associated with historical SNe with
estimated ages of a few hundred years or less have been found, namely, 
SNR~4449-1 in NGC~4449 and SNR~B12-174a in M83 (see Table~\ref{fluxes}).  

This situation, together with the small sample size of young, $\leq$ 1000 yr
old Galactic CCSN remnants (i.e., the Crab, 3C58, and Cas A), makes further
study of the 78 yr old SN~1941C across multi-wavelengths especially valuable
for investigating the SN-SNR connection.  This could include {\sl HST}
multiband imaging of the explosion site and stellar association, deep {\sl
Chandra} imaging for X-ray spectral analysis, an investigation of its likely
late-time radio emission, as well as higher S/N optical spectra.


\acknowledgements

We thank Bill Blair, Frank Winkler, and Dan Milisavljevic for sharing
late-time spectral data with us, Dan Patnaude for help with the X-ray analysis
of SN 1986E, David H.\ Roberts for his helpful discussions of VLA observations
of NGC~4136, an anonymous referee for several helpful comments and corrections, and
Eric Galayda and the MDM staff for assistance with our observations. R.A.F.\
and K.E.W\ acknowledge support from STScI Guest Observer Programs 15337 and
15515.  K.E.W.\ also acknowledges support from Dartmouth's Guarini School of
Graduate and Advanced Studies, and the Chandra X-ray Center under CXC grant
GO7-18050X.  This work is part of R.A.F's Archangel III Research Program.
Finally, we would like to acknowledge the extensive research on SNe done by
Milton Humason and Rudolph Minkowski which included the spectroscopic
classification of SN~1941C, the subject of this paper.

\facilities{Hiltner (OSMOS), Sloan}

\software{PYRAF \citep{pyrafcite}, AstroPy \citep{AstropyA,AstropyB}, ds9 \citep{ds9cite}, L.A.\ Cosmic \citep{vanDokkum2001}}


\end{document}